# GPU Parallelization Strategies for Forward and Backward Propagation in Shallow Neural Networks: A CUDA-Based Comparative Study

Nadine Bousdjira, Sarah Hasnaoui, Amel Sadoun, Fatma Salhi, Rania Zitouni
École Nationale Supérieure d'Informatique (ESI)
High Performance Computing (HPC)
Instructor: Mrs. Amina Haichour
Academic Year 2025–2026

***Abstract*—This paper compares two GPU parallelization strategies for training shallow neural networks using CUDA. We analyze a baseline implementation that uses naive matrix multiplication and identify its main bottlenecks: redundant global memory accesses and poor memory coalescing. To address these issues, we propose an optimized strategy that combines tiled matrix multiplication with shared memory, matrix transposition for coalesced access, and kernel fusion to reduce memory transfers. Experiments on an NVIDIA GPU show that our optimized approach achieves approximately 1.41× speedup on large datasets compared to the baseline. Training time for the largest dataset decreased from 21.00s to 14.82s. Our results demonstrate that memory-aware optimizations can significantly improve neural network training performance on GPUs.**

## I. Introduction

Neural networks have become widely used in many applications, but training them requires significant computational resources. Even shallow neural networks can take considerable time to train, especially with large datasets. GPUs offer a solution to this problem due to their ability to perform many calculations in parallel. However, simply running code on a GPU does not guarantee good performance. We need to carefully design our parallelization strategy to take advantage of the GPU architecture.

CUDA (Compute Unified Device Architecture) is NVIDIA's platform for programming GPUs. It gives us direct control over how we organize threads, manage memory, and execute computations. This low-level control allows us to optimize our code for specific tasks like neural network training.

### A. Motivation

CPUs are limited in how many operations they can perform simultaneously. Neural network training involves many matrix operations that can be computed independently, making them good candidates for parallel execution. The forward and backward propagation algorithms are particularly well-suited for GPUs because they consist mainly of matrix multiplications that can be parallelized.

However, achieving good performance on GPUs requires more than just parallel execution. We must consider how threads access memory, how data is reused, and how to minimize synchronization overhead. A naive implementation may not fully utilize the GPU's capabilities.

### B. Problem Statement

This study focuses on efficiently parallelizing shallow neural network training on GPUs. We want to understand what makes a parallelization strategy perform well and what causes performance bottlenecks.

### C. Contributions

Our work includes:

1) Analysis of a baseline CUDA implementation, identifying where it performs poorly and why.
2) Design and implementation of an improved parallelization strategy.
3) Experimental comparison of both strategies using different dataset sizes and network configurations.
4) Practical insights for implementing neural network training on GPUs.

### D. Paper Organization

The rest of this paper is organized as follows: Section 2 covers background on neural networks and GPU programming. Section 3 describes our network architecture. Section 4 analyzes the baseline strategy. Section 5 presents our optimized approach. Section 6 shows experimental results. Section 7 discusses our findings, and Section 8 concludes.

## II. Background and Related Work

### A. Shallow Neural Networks

A shallow neural network has one hidden layer between the input and output layers. Our network has 32 input neurons, 256 hidden neurons, and 1 output neuron. We use ReLU activation and Mean Squared Error (MSE) loss for regression. Training uses mini-batch Stochastic Gradient Descent (SGD).

Forward propagation consists of two matrix multiplications with ReLU activation in between. For a batch of input samples $X \in \mathbb{R}^{n \times 32}$:

$$Z_1 = XW_1 \quad (1)$$

$$H = \mathrm{ReLU}(Z_1) \tag{2}$$

$$Y_{\text{pred}} = HW_2 \tag{3}$$

where $W_1 \in \mathbb{R}^{32\times256}$ and $W_2 \in \mathbb{R}^{256\times1}$ are the weight matrices.

Backpropagation computes gradients using the chain rule. With MSE loss:

$$L = \frac{1}{n}\sum_{i=1} \left(Y^{(i)} - Y^{(i)}_{\text{pred}}\right)^2 \tag{4}$$

The gradients are:

$$dZ_2 = \frac{2}{n}(Y_{\text{pred}} - Y) \tag{5}$$

$$dW_2 = H^T dZ_2 \tag{6}$$

$$dZ_1 = (dZ_2 W_2^T) \odot \mathrm{ReLU}'(Z_1) \tag{7}$$

$$dW_1 = X^T dZ_1 \tag{8}$$

Weight updates use gradient descent:

$$W^{(l)} = W^{(l)} - \alpha dW^{(l)} \tag{9}$$

Most of the computation time comes from matrix multiplications in both forward and backward passes.

### B. GPU Architecture and CUDA Programming

GPUs contain multiple Streaming Multiprocessors (SMs) that can run thousands of threads concurrently. CUDA organizes threads hierarchically:

- **Grid**: Collection of thread blocks
- **Block**: Group of threads that can share data and synchronize
- **Thread**: Individual execution unit

GPUs have several memory types with different speeds and sizes:

- **Global Memory**: Large (several GB) but slow (hundreds of cycles latency)
- **Shared Memory**: Small (tens of KB per block) but fast (few cycles latency)
- **Registers**: Very fast but limited per thread
- **L1/L2 Cache**: Automatic hardware caching

Good GPU performance requires coalesced memory access, where adjacent threads access adjacent memory locations. This allows the GPU to combine multiple accesses into fewer memory transactions [1].

## III. Neural Network Architecture

### A. Network Structure

We use a simple fully connected network with one hidden layer:

- Input layer: 32 neurons
- Hidden layer: 256 neurons
- Output layer: 1 neuron
- Activation: ReLU
- Loss function: MSE
- Batch size: 256
- Learning rate: 0.002
- Epochs: 100

The network has no bias terms, which simplifies implementation. Total parameters:

$$(32 \times 256) + (256 \times 1) = 8,448 \tag{10}$$

### B. Forward Propagation

For input batch $X \in \mathbb{R}^{n\times32}$, we compute:

$$Z_1 = XW_1, \quad W_1 \in \mathbb{R}^{32\times256} \tag{11}$$

$$H = \mathrm{ReLU}(Z_1), \quad \mathrm{ReLU}(x) = \max(0, x) \tag{12}$$

$$Y_{\text{pred}} = HW_2, \quad W_2 \in \mathbb{R}^{256\times1} \tag{13}$$

Computational cost is dominated by the two matrix multiplications: $O(n \cdot 32 \cdot 256)$ and $O(n \cdot 256 \cdot 1)$.

### C. Backward Propagation

We minimize MSE loss:

$$L = \frac{1}{n}\sum_{i=1}^{n} \left(Y^{(i)}_{\text{pred}} - Y^{(i)}\right)^2 \tag{14}$$

Gradients are computed as:

$$dZ_2 = \frac{2}{n}(Y_{\text{pred}} - Y) \tag{15}$$

$$dW_2 = H^T dZ_2 \tag{16}$$

$$dZ_1 = (dZ_2 W_2^T) \odot \mathrm{ReLU}'(Z_1) \tag{17}$$

$$dW_1 = X^T dZ_1 \tag{18}$$

where $\mathrm{ReLU}'(x) = 1$ if $x > 0$, otherwise $0$.

Weight update:

$$W^{(l)} = W^{(l)} - \alpha dW^{(l)}, \quad \alpha = 0.002 \tag{19}$$

Like forward propagation, backward propagation is dominated by matrix multiplications.

## IV. Baseline Parallelization Strategy

### A. Overall Approach

The baseline strategy parallelizes only the matrix multiplication operations ($C = A \times B$). Everything else (activation functions, loss calculation, control logic) runs on the CPU or as separate simple GPU kernels. This is a straightforward approach that prioritizes simplicity over optimization.

Each matrix multiplication needed during training launches a separate GPU kernel. While this makes the code easy to understand, it misses opportunities for optimization.

### B. Thread Organization

*1) Parallelization Granularity:* The baseline uses a 2D grid of 2D blocks:

- **Granularity:** One thread per output element
- **Block size:** $16 \times 16$ threads (256 threads per block)
- **Grid size:** $\lceil M/16 \rceil \times \lceil N/16 \rceil$ blocks for $M \times N$ output
- **Mapping:** Thread at position $(row, col)$ computes $C[row][col]$

*2) Computation:* For $C = A \times B$ where $A \in \mathbb{R}^{M \times K}$ and $B \in \mathbb{R}^{K \times N}$, each thread computes:

$$C[i][j] = \sum_{k=0}^{K-1} A[i][k] \times B[k][j] \quad (20)$$

All matrices are stored in global memory. There is no shared memory usage, no tiling, and no data reuse strategy. Each thread reads directly from global memory for every element it needs.

### C. Memory Access Pattern

*1) Memory Hierarchy:* The baseline only uses global memory:

- **Global Memory:** All matrices ($A$, $B$, $C$)
- **Shared Memory:** Not used
- **Registers:** Only for the accumulator variable
- **Caching:** Relies on automatic L1/L2 caches

*2) Access Pattern:* Threads access matrix $A$ row-wise:

$$A[row][0], A[row][1], \ldots, A[row][K-1] \quad (21)$$

Threads access matrix $B$ column-wise:

$$B[0][col], B[1][col], \ldots, B[K-1][col] \quad (22)$$

Row-wise access to $A$ is good because adjacent threads access adjacent memory. But column-wise access to $B$ is problematic. Since matrices are stored row-major, accessing columns means reading elements that are far apart in memory (strided access). This prevents memory coalescing and wastes bandwidth.

### D. Computational Cost

For batch size $n = 256$, the work per epoch includes:

**Forward Pass:**

- $Z_1 = XW_1$: $n \times 32 \times 256 = 2,097,152$ FLOPs
- $Y_{pred} = HW_2$: $n \times 256 \times 1 = 65,536$ FLOPs

**Backward Pass:**

- $dW_2 = H^T dZ_2$: $256 \times n \times 1 = 65,536$ FLOPs
- $dZ_1 = dZ_2 W_2^T$: $n \times 1 \times 256 = 65,536$ FLOPs
- $dW_1 = X^T dZ_1$: $32 \times n \times 256 = 2,097,152$ FLOPs

Total: approximately 4.4 million FLOPs per batch.

### E. Performance Bottlenecks

After analyzing the baseline, we identified several major problems:

*1) Redundant Global Memory Access:* The biggest issue is that the same data is read from global memory many times. For an $N \times N$ matrix multiplication:

- Each of $N^2$ threads reads $N$ elements from $A$ and $N$ elements from $B$
- Total reads: $2N^3$ memory transactions
- For our $256 \times 256$ multiplication, each matrix element is fetched 256 times by different threads

This redundancy saturates memory bandwidth (around 900 GB/s on modern GPUs), leaving compute cores waiting for data.

*2) Uncoalesced Memory Access:* Matrix $A$ is accessed row-wise, which gives coalesced access (good). But matrix $B$ is accessed column-wise, which creates strided access (bad). In row-major storage:

- Adjacent threads access $B[k][col]$, $B[k][col+1]$, etc.
- These addresses differ by $K$ elements (the row width), not 1
- The GPU cannot combine these into efficient transactions
- This can reduce effective bandwidth by up to 75%

*3) Memory-Bound Performance:* Each thread does 256 multiply-add operations but must read 512 elements from global memory. The ratio of computation to memory access is low. This means the implementation is memory-bound: cores spend more time waiting for data than computing.

*4) No Data Reuse:* The baseline has no strategy for reusing data:

- No shared memory for threads in a block to share data
- No tiling to break large multiplications into cache-friendly pieces
- Only relies on automatic L1/L2 caching, which is insufficient

*5) Kernel Launch Overhead:* Each operation (matrix multiply, activation) launches a separate kernel:

- Intermediate results are written to and read from global memory
- Each launch has synchronization overhead (microseconds)
- Over 100 epochs, this adds up to significant wasted time

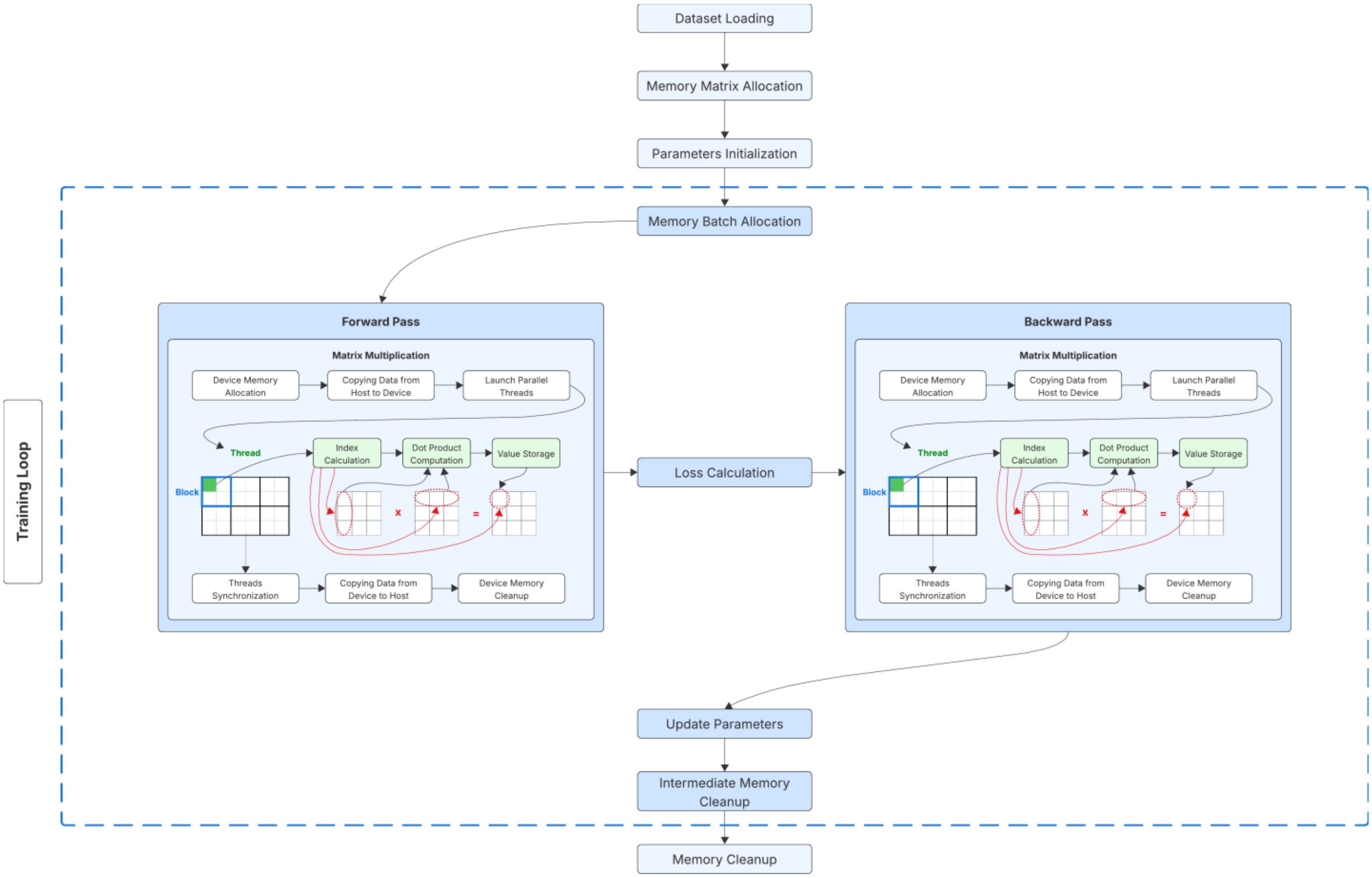

Figure 1: Baseline strategy: each thread computes one output element using only global memory. No shared memory or tiling is used.

These bottlenecks show that the baseline is limited mainly by memory performance, not computation. This suggests that memory optimizations could significantly improve performance.

## V. Proposed Alternative Parallelization Strategy

To address the baseline's limitations, we implemented three memory optimization techniques: tiled matrix multiplication with shared memory, matrix transposition for better memory access, and kernel fusion to reduce memory transfers.

### A. Tiled Matrix Multiplication with Shared Memory

*1) Motivation:* In the baseline, each thread reads a full row from $A$ and a full column from $B$ from global memory to compute one output element. Since many threads need the same data, this causes massive redundancy. The same data is fetched hundreds of times. Global memory is slow (hundreds of cycles latency), so cores wait idle for data most of the time.

*2) Method:* We use shared memory, which is significantly faster than global memory (latency of only a few cycles). The idea is to divide the matrices into small tiles of size $TILE_SIZE \times TILE_SIZE$ and process them in phases:

1) Threads within a block cooperatively load one tile from matrix $A$ and one tile from matrix $B$ into shared memory.
2) All threads synchronize using `__syncthreads()` to ensure that the data is fully loaded.
3) Each thread computes a partial result using the data stored in fast shared memory.
4) Threads synchronize again before loading the next tile.

This reduces global memory accesses from $2N^3$ to approximately $2N^2 N$ for an $N \times N$ multiplication. Each element is loaded from global memory only once per block, not once per thread.

### B. Matrix Transposition for Coalesced Access

After implementing tiling, we still had inefficient access to matrix $B$ because of the column-wise access pattern.

*1) Motivation:* Reading rows from $A$ is efficient because addresses are consecutive $(i, i + 1, i + 2, \ldots)$. The GPU can fetch this in one transaction (coalesced access). But reading columns from $B$ in row-major storage means addresses are spread out $(i, i + width, i + 2 \cdot width, \ldots)$. This strided access is slow because it requires many separate transactions.

*2) Method:* We added a transposition step for matrix $B$. By computing $B^T$ first, the columns of $B$ become rows of $B^T$.

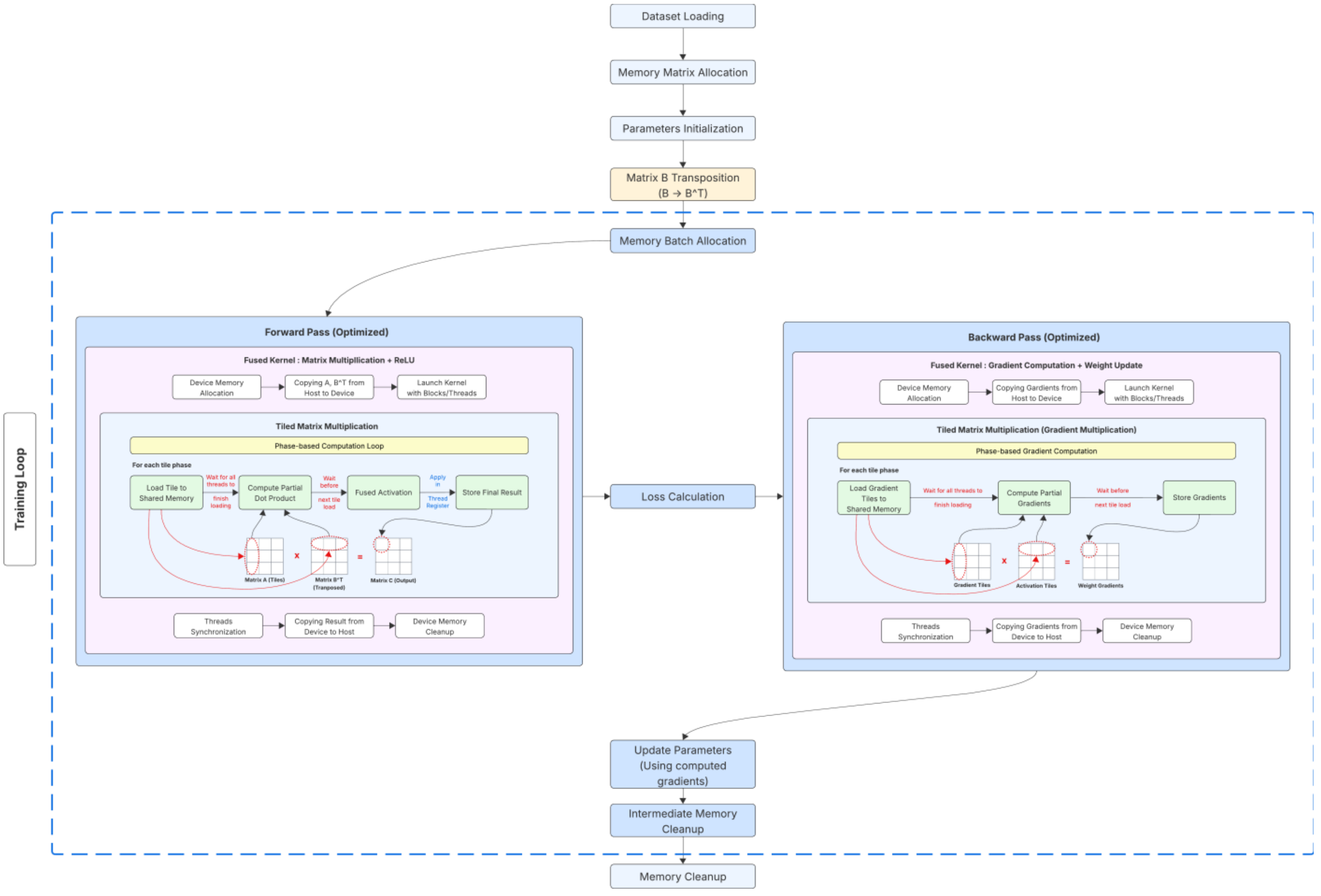

Figure 2: Optimized strategy using tiled multiplication with shared memory, matrix transposition for coalesced access, and kernel fusion.

Now when we compute $C = A \times B^T$ , both matrices are read row-wise, giving fully coalesced access.

The kernel accesses the second matrix differently:

- **Before:** val = B[k * width + col] (strided, slow)
- **After:** val = B_transposed[col * width + k] (coalesced, fast)

Transposing has a small cost, but it happens only once per epoch while the weights are used for many batches, so the benefit outweighs the cost.

### C. Kernel Fusion

*1) Motivation:* In the baseline, matrix multiplication and activation run as separate kernels. The multiplication writes results to global memory, then the activation kernel reads them back. Writing and reading intermediate results is expensive and wastes bandwidth.

*2) Method:* We combined these operations into one kernel. After computing the dot product, the thread immediately applies ReLU while the data is still in registers:

$$Output = \text{ReLU}\sum (A_{ik} \times B_{kj}) \tag{23}$$

This eliminates one complete round-trip to global memory, reducing execution time.

## VI. Experimental Evaluation

### A. Experimental Setup

We ran all experiments on an NVIDIA Tesla T4 GPU (Compute Capability 7.5, around 1550 MHz clock, 15 GB memory) using CUDA 13.0. We compared:

- **Baseline:** Global memory only
- **Optimized:** Tiling + shared memory + transposition + fusion

We measured:

- Execution time (forward + backward pass)
- Speedup vs baseline
- Scalability with dataset size
- Impact of network size

All experiments used the same conditions:

- 100 epochs
- Batch size = 256
- Same learning rate and dataset
- Same hardware

## B. Impact of Dataset Size

We tested three dataset sizes:

- **Small:** 256 samples
- **Medium:** 2,560 samples
- **Large:** 25,600 samples

Network architecture was fixed (32 input, 256 hidden, 1 output) to isolate the effect of dataset size.

*1) Execution Time:* Table I shows total training time for 100 epochs.

Table I: Training Time Comparison (100 Epochs)

| **Dataset** (Samples) | **Baseline** (Seconds) | **Optimized** (Seconds) | **Speedup** (Ratio) |
|---|---|---|---|
| Small (256) | 0.306 s | 0.248 s | 1.23× |
| Medium (2,560) | 2.000 s | 1.561 s | 1.28× |
| Large (25,600) | 21.002 s | 14.823 s | **1.41×** |

The optimized strategy is consistently faster for all dataset sizes.

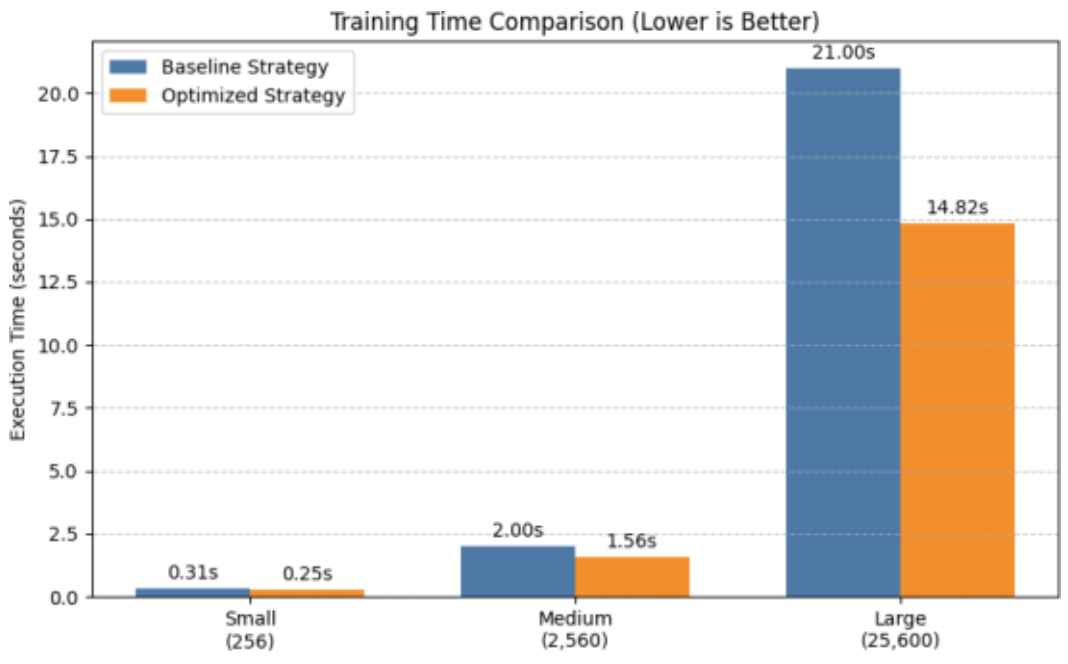

Figure 3: Training time comparison for different dataset sizes.

*2) Scalability:* Speedup increases with dataset size:

- Small: 1.23×
- Medium: 1.28×
- Large: 1.41×

For small datasets, kernel launch overhead and CPU control logic take up a larger portion of total time. For large datasets, the GPU is better utilized and our memory optimizations have more impact. Tiling reduces global memory traffic and shared memory helps hide latency, which explains the increasing speedup.

This confirms that our optimizations work better when the workload is large enough to fully use the GPU.

## C. Impact of Network Size

We fixed the dataset at 2,560 samples and varied hidden neurons: 32, 64, 128, 256, 512, 1024. More neurons means more parameters and more computation.

Training time increases almost linearly with neuron count, which makes sense because matrix multiplication complexity scales as $O(batch_size \times input_size \times hidden_size)$. Throughput decreases with larger networks because more neurons create more memory traffic and computation.

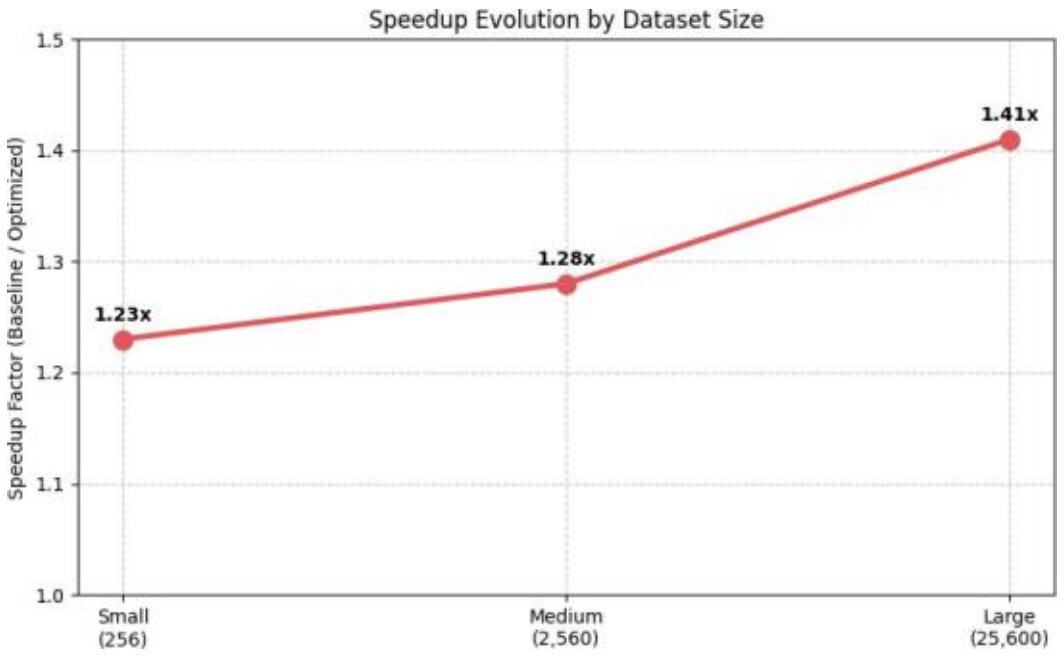

Figure 4: Speedup evolution with dataset size.

Table II: Training Time vs Hidden Layer Size

| **Hidden Neurons** | **Time (s)** | **Speedup vs 32** |
|---|---|---|
| 32 | 0.810 | 1.00× |
| 64 | 0.921 | 0.88× |
| 128 | 1.107 | 0.73× |
| 256 | 1.551 | 0.52× |
| 512 | 2.743 | 0.30× |
| 1024 | 4.645 | 0.17× |

# VII. Discussion

## A. Interpretation of Results

The experiments clearly show that the optimized version performs better, especially for large datasets. This confirms that the baseline was limited mainly by memory bandwidth, not by computation.

Shared memory reduces global memory accesses because each tile is loaded once and reused by all threads in the block. This improves data reuse and increases the ratio of computation to memory access. The padding technique (`tile[16][17]`) avoids bank conflicts in shared memory. Without padding, some accesses would serialize and hurt performance.

The neuron size experiments show that training time grows almost linearly with hidden layer size, which is expected from the matrix multiplication complexity. The GPU performs better when the workload is large enough to keep all cores busy.

## B. Limitations

Our study has some limitations:

- We only tested a single hidden layer network. Deeper networks might show different performance characteristics.
- Some training logic remains on the CPU, adding kernel launch overhead.
- We only measured execution time. We did not analyze occupancy, memory bandwidth, or FLOPS using profiling tools like Nsight.
- The neuron experiments used only one dataset size. Combining both variations would give a more complete picture.

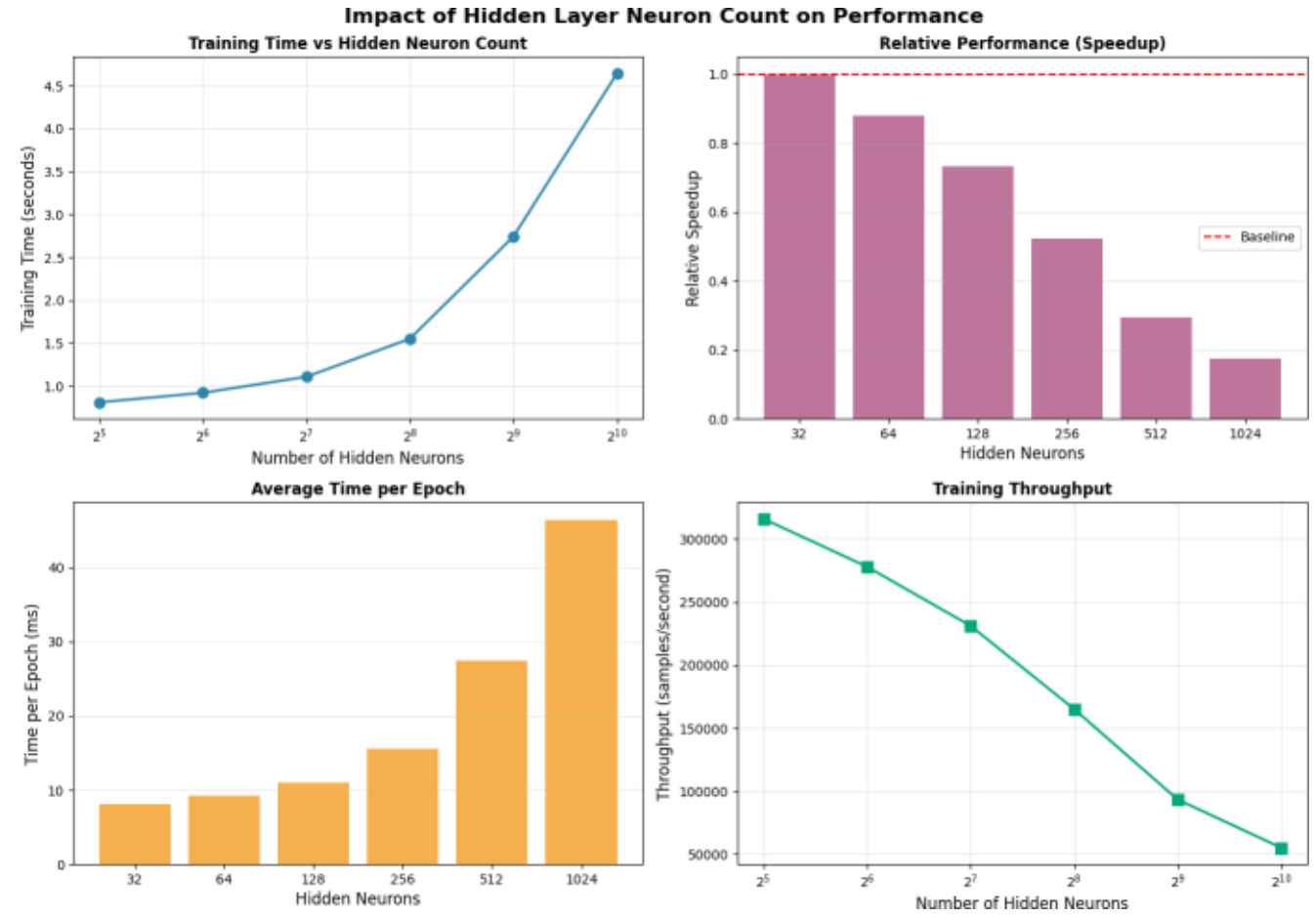

Figure 5: Impact of hidden layer size on training performance.

Future work could include detailed profiling with Nsight, testing deeper networks, and moving more control logic to the GPU to reduce overhead.

## VIII. Conclusion

In this study, we optimized shallow neural network training on CUDA by addressing memory bottlenecks. We identified global memory redundancy and uncoalesced access as the main problems in the baseline implementation. Our solution combined tiled matrix multiplication, shared memory padding, and kernel fusion, achieving approximately 30% faster training on large datasets (1.41 $\times$speedup).

This project showed us that GPU performance depends not just on parallel execution, but on efficient memory management. The number of operations matters less than how efficiently we access and reuse data. Understanding the GPU memory hierarchy and designing around it is crucial for high performance computing.

## Acknowledgments

We thank our HPC teacher Mrs. Amina HAICHOUR and the École nationale Supérieure d'Informatique (ESI) for providing the resources and guidance for this research project.

## References

[1] NVIDIA Corporation, "CUDA C Programming Guide," 2024. [Online]. Available: https://docs.nvidia.com/cuda/

[2] I. Goodfellow, Y. Bengio, and A. Courville, *Deep Learning*. Cambridge, MA, USA: MIT Press, 2016.

[3] J. Sanders and E. Kandrot, *CUDA by Example: An Introduction to General-Purpose GPU Programming*. Boston, MA, USA: Addison-Wesley Professional, 2010.

[4] S. Chetlur, C. Woolley, P. Vandermersch, J. Cohen, J. Tran, B. Catanzaro, and E. Shelhamer, "cuDNN: Efficient primitives for deep learning," *arXiv preprint arXiv:1410.0759*, 2014.

[5] Y. Jia, E. Shelhamer, J. Donahue, S. Karayev, J. Long, R. Girshick, S. Guadarrama, and T. Darrell, "Caffe: Convolutional architecture for fast feature embedding," in *Proc. 22nd ACM Int. Conf. Multimedia*, Orlando, FL, USA, Nov. 2014, pp. 675–678.

[6] V. Volkov and J. W. Demmel, "Benchmarking GPUs to tune dense linear algebra," in *Proc. 2008 ACM/IEEE Conf. Supercomputing (SC'08)*, Austin, TX, USA, Nov. 2008, pp. 1–11.

[7] S. Li, A. Mishra, J. J. Doherty, M. Beadon, and S. Cadambi, "Neural network acceleration study with ReRAM: Opportunities and challenges," in *Proc. IEEE Int. Symp. Performance Anal. Syst. Softw. (ISPASS)*, Uppsala, Sweden, Apr. 2016, pp. 197–198.

[8] M. Abadi *et al.*, "TensorFlow: A system for large-scale machine learning," in *Proc. 12th USENIX Symp. Operating Syst. Design Implementation (OSDI)*, Savannah, GA, USA, Nov. 2016, pp. 265–283.